# Resonant Band Engineering of Ferroelectric Tunnel Junctions

Jing Su[1], Xingwen Zheng[1], Zheng Wen[2], Tao Li[3], Shijie Xie[1], Karin M. Rabe[4], Xiaohui Liu[1*], and Evgeny Y. Tsymbal[5**]

[1] *School of Physics, State Key Laboratory of Crystal Materials, Shandong University, Ji'nan 250100, China*
[2] *College of Physics and Center for Marine Observation and Communications, Qingdao University, Qingdao 266071, China*
[3] *Center for Spintronics and Quantum Systems, State Key Laboratory for Mechanical Behavior of Materials, Department of Materials Science and Engineering, Xi'an Jiaotong University, Xi'an, Shanxi 710049, China*
[4] *Department of Physics and Astronomy, Rutgers University, Piscataway, New Jersey 08854, USA*
[5] *Department of Physics and Astronomy, University of Nebraska, Lincoln, Nebraska 68588-0299, USA*

We propose energy band engineering to enhance tunneling electroresistance (TER) in ferroelectric tunnel junctions (FTJs). We predict that an ultrathin dielectric layer with a smaller band gap, embedded into a ferroelectric barrier layer, acts as a switch controlling high and low conductance states of an FTJ depending on polarization orientation. Using first-principles modeling based on density functional theory, we investigate this phenomenon for a prototypical $SrRuO_3/BaTiO_3/SrRuO_3$ FTJ with a $BaSnO_3$ monolayer embedded in the $BaTiO_3$ barrier. We show that in such a composite-barrier FTJ, ferroelectric polarization of $BaTiO_3$ shifts the conduction band minimum of the $BaSnO_3$ monolayer above or below the Fermi energy depending on polarization orientation. The resulting switching between direct and resonant tunneling leads to a TER effect with a giant ON/OFF conductance ratio. The proposed resonant band engineering of FTJs can serve as a viable tool to enhance their performance useful for device application.

A ferroelectric tunnel junction (FTJ) is a functional electronic device with electrical resistance being controlled by ferroelectric polarization [1]. A typical FTJ is composed of two conducting electrodes separated by a nanometer-thick ferroelectric layer, which serves as a tunnel barrier. A figure of merit of an FTJ is tunneling electroresistance (TER)—a resistance change resulting from polarization reversal of the ferroelectric barrier layer [2,3]. Such polarization switching allows the control of two non-volatile resistance states of an FTJ (low and high) which can be employed in random access memories and other electronic devices [4]. Enhancing the magnitude of TER is beneficial for device application of FTJs.

There are several physical mechanisms responsible for the TER effect [5]. Most of them involve a modulation of the effective tunneling barrier encountered by transport electrons and driven by reversal of ferroelectric polarization [6]. Different microscopic processes control the tunneling barrier in FTJs, involving those at the interfaces, within the electrodes, as well as in the ferroelectric layer itself. An important perquisite for obtaining a large TER is asymmetry of the FTJ in respect to its electronic and atomic structure. Specifically, it has been demonstrated that sizable TER can be obtained using dissimilar electrodes [7-13], a composite barrier layer [14-17], and interface engineering [18-22]. Interesting physical phenomena have been predicted and demonstrated in FTJs, including ferroelectric-induced magnetic interface phase transition [23-25], tunneling barrier metallization [26, 27], defect-controlled [28-32] and bias-modulated [33, 34] transport, and tunneling across an in-plane domain wall [35, 36].

Despite this notable progress, further improvements in the FTJ performance are required to meet industry demands. Nowadays, due to advances in thin-film deposition techniques, growth of thin-film heterostructures can be controlled with the atomic scale precision. This allows tuning the atomic structure of FTJs within a single atomic layer to achieve the required electronic and transport properties. For example, using a layer-by-layer growth, δ-doping can be realized to improve the performance of FTJs and associated electronic devices.

In this work, we propose that the transport properties of FTJs can be significantly enhanced by inserting an ultrathin layer of a dielectric with a relatively smaller band gap in the ferroelectric barrier. Such a dielectric layer produces resonant states in the barrier energy gap which can be controlled by ferroelectric polarization. Polarization reversal shifts the conduction band of the dielectric layer up and down with respect to the Fermi level, which results in the switching of the transport regime between direct and resonant tunneling. Such a resonant-band control strongly enhances an ON/OFF conductance ratio of FTJs and provides a practical tool to engineer their electronic and transport properties to meet industry requirements.

To explore this mechanism of TER, we consider a prototypical FTJ which consists of $SrRuO_3$ electrodes and $BaTiO_3$ ferroelectric barrier with one $TiO_2$ atomic layer being substituted with $SnO_2$. The substitution of $SnO_2$ for $TiO_2$ can be thought as insertion of one-unit-cell-thick $BaSnO_3$ layer instead of that of $BaTiO_3$. $BaSnO_3$ has been predicted to exhibit a conduction band offset with respect to $BaTiO_3$ as large as $-1.13$ eV [37]. In addition, $BaSnO_3$ has a similar lattice constant to $BaTiO_3$ and both $BaTiO_3$ and $BaSnO_3$ can grow epitaxially on $SrTiO_3$ [38]. Thus, if such a one-unit-cell-thick $BaSnO_3$ layer is inserted in the tunneling barrier, the conduction band minimum (CBM) of this layer will be shifted down with respect to the CBM of $BaTiO_3$, affecting the transport mechanism. When the polarization of $BaTiO_3$ is switched between two polarization states, the CBM of $BaSnO_3$ will be pushed up and down in energy, above and below the Fermi level, so that the conduction mechanism changes from direct to resonant tunneling. This will lead to a large change in resistance of the FTJ and thus a sizable TER effect.

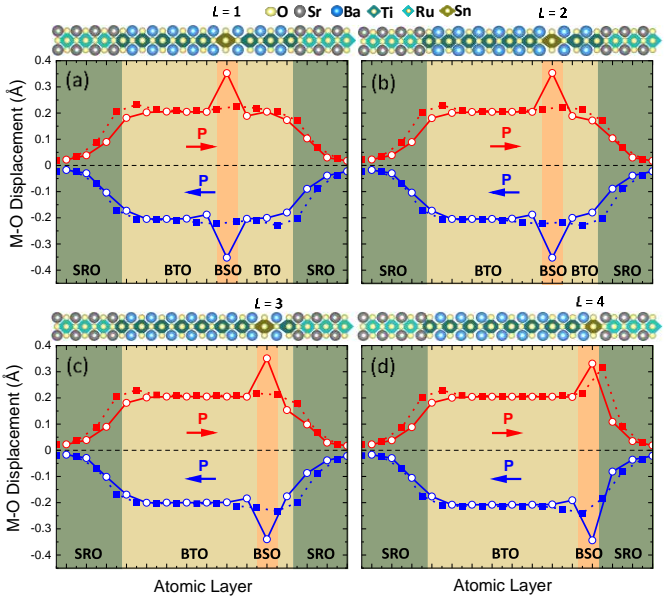

FIG. 1. Calculated relative polar displacement between cation (M) and anion (O) on each $AO$ ($A$ = Ba or Sr) and $BO_2$ ($B$ = Ti, Sn, or Ru) layer across SrRuO$_3$/BaTi(Sn)O$_3$/SrRuO$_3$ supercells (top panels) with a SnO$_2$ monolayer placed at four different positions numbered by $L = 1$ (a), $L = 2$ (b), $L = 3$ (c), and $L = 4$ (d) as indicated by orange vertical bars. Positive (negative) values of the displacement shown in red (blue) correspond to polarization pointing to the right (left). Open (solid) symbols denote Ti-O, Sn-O, and Ru-O (Ba-O and Sr-O) displacements.

To confirm these expectations, we perform first-principles calculations, as described in Supplemental Material [39]. In the calculations, we use a supercell constructed of 8.5 unit cells of BaTiO$_3$ with one atomic layer of TiO$_2$ substituted by SnO$_2$ and 6.5 unit cells of SrRuO$_3$. The SnO$_2$ is placed at four different positions numbered by $L$, as indicated on top panels of Figs. 1 (a-d). To simulate coherent epitaxial growth of the structure on an SrTiO$_3$ substrate in experiment, we constrain the in-plane lattice constant of the supercell to the calculated lattice constant of cubic SrTiO$_3$, $a = 3.94$ Å. This constraint imposes an in-plane strain of about $-1.5\%$ on BaTiO$_3$ and stabilizes BaTiO$_3$ in the $P4mm$ tetragonal phase. Under this constraint, the calculated direct bands gap of BaTiO$_3$ and BaSnO$_3$ are found to be 2.5 eV and 1.2 eV, respectively. The smaller band gap of BaSnO$_3$ leads to the appearance of quantum-well states in the band gap of BaTiO$_3$, when Sn is substituted for Ti in a monolayer or two monolayers of BaTiO$_3$ [39].

We find that ferroelectric polarization is switchable in all FTJ structures independent of the position of the inserted SnO$_2$ layer. Figs. 1 (a-d) show the respective cation-anion displacement at each $AO$ ($A$ = Ba or Sr) and $BO_2$ ($B$ = Ti, Sn, or Ru) atomic layer in these FTJs for polarization pointing left (blue curves) and right (red curves). The Ti-O and Ba-O displacements in the barrier layer are consistent with the previous calculations for pure BaTiO$_3$ [40], reflecting a bulk-like polarization of BaTiO$_3$ independent of the position of the SnO$_2$ layer. It is notable that the SnO$_2$ layer itself behaves as a strong dipole which is signified by a large Sn-O displacement. The presence of the Ru-O and Sr-O displacement, decaying away from the interface into the SrRuO$_3$ electrode, reflects the effect of ionic screening [41].

Fig. 2 shows the calculated local density of states (LDOS) projected on TiO$_2$ and SnO$_2$ layers. It is seen that there is band bending across the barrier due to a depolarizing field, whose direction depends on polarization orientation. For polarization pointing right (Figs. 2 (a-d)), the CBM on the SnO$_2$ layer quickly approaches the Fermi energy $E_F$ when this layer is moved from the middle of the barrier layer ($L = 1$) to the interface ($L = 4$). When SnO$_2$ is in layer $L = 2$, i.e. two TiO$_2$ layers away from the right interface (Fig. 2 (b)), the SnO$_2$ LDOS touches $E_F$, and when it is in layer $L = 3$ (Fig. 2 (c)), bottom of the SnO$_2$ LDOS lies about 0.2 eV below $E_F$. In contrast, for polarization pointing left (Figs. 2 (e-h)), the CBM on the SnO$_2$ layer lies well above the Fermi level due to the asymmetric placement of this layer closer to the right interface.

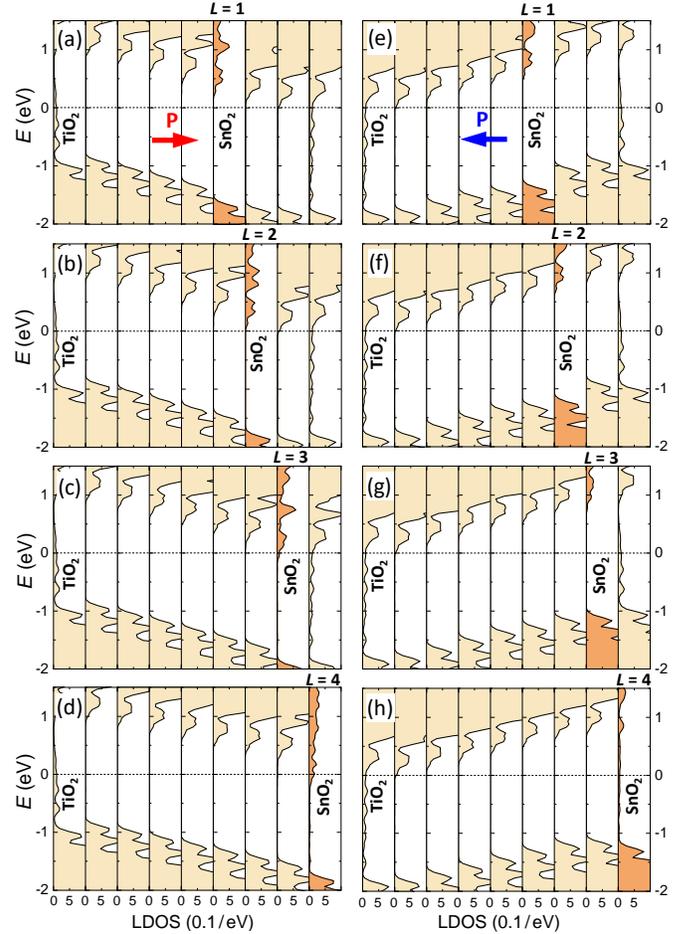

FIG. 2. Local densities of states (LDOS) as a function of energy $E$ on each $BO_2$ ($B$ = Ti, Sn) atomic layer within a composite BaTi(Sn)O$_3$ barrier for four FTJ structures with SnO$_2$ layer at $L = 1$ (a, e), $L = 2$ (b, f), $L = 3$ (c, g), and $L = 4$ (d, h) for polarization pointing right (a-d) and left (e-h). Horizontal dashed lines indicate the Fermi energy.

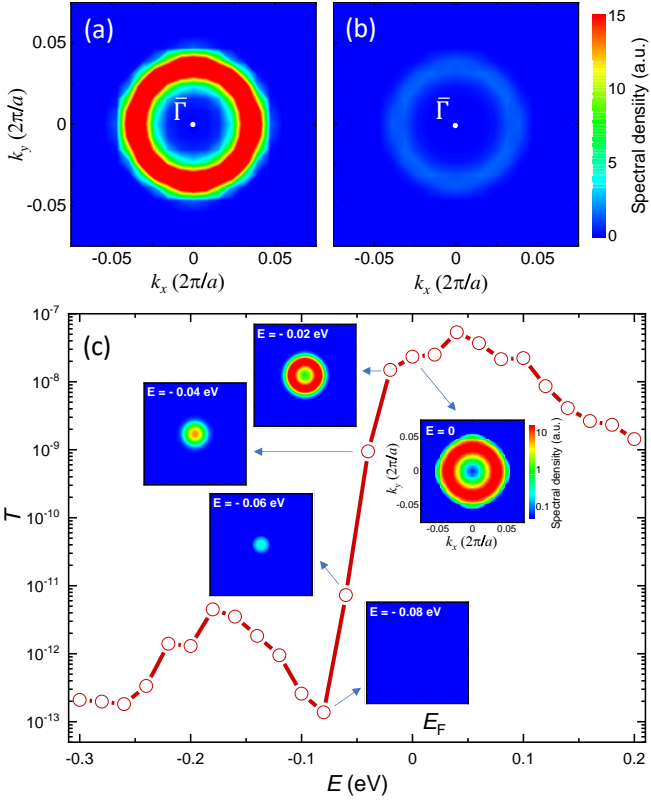

FIG. 3. Spectral density around the $\bar{\Gamma}$ point ($k_\parallel = 0$) projected onto the $SnO_2$ (a) and interfacial $TiO_2$ (b) layers at the Fermi energy for $SrRuO_3/BaTiO_3(BaSnO_3)/SrRuO_3$ FTJ with $SnO_2$ layer at $L = 3$ and ferroelectric polarization pointing to the right. (c) Transmission $T$ per unit-cell area of the FTJ as a function of electron energy $E$. The Fermi energy $E_F$ is at zero. Insets display spectral densities in the logarithmic scale at $E = 0, -0.02, -0.04, -0.06$, and $-0.08$ eV.

The appearance of the $SnO_2$ electronic states at the Fermi energy provides a resonant channel for conductance, which is responsible for a large TER effect discussed below. The presence of resonant states is evident from Fig. 3 (a) showing the spectral density (SD) on the $SnO_2$ layer around the $\bar{\Gamma}$ point ($k_\parallel = 0$) at $E_F$ for FTJ with $SnO_2$ placed at $L = 3$ (corresponding to the LDOS in Fig. 2 (c)). The high SD at a ring of radius $k_\parallel \approx 0.035 \frac{2\pi}{a}$ indicates the presence of a 2D free-electron-like band localized within the $SnO_2$ layer. While the adjacent interfacial $TiO_2$ layer at the right interface is also metalized (as follows from the non-zero LDOS at $E_F$ in Fig. 2 (c)), it has a very small SD in the area around the $\bar{\Gamma}$ point where the $SnO_2$ SD is sizable (Fig. 3 (b)). In fact, this $TiO_2$ SD represents $SnO_2$-induced gap states, and the $TiO_2$-layer metallization largely results from $\boldsymbol{k}_\parallel$ lying beyond the area shown in Fig. 3 (b). Therefore, the interfacial $TiO_2$ layer provides an effective barrier for tunneling electrons with transverse wave vectors $\boldsymbol{k}_\parallel$ corresponding to the quantum-well states on $SnO_2$.

When the electron energy $E$ is shifted down below $E_F$, the $SnO_2$ spectral density shrinks to a ring of a smaller radius (insets in Fig. 3 (c)) and then collapses to a point (inset in Fig. 3 (c) for $E = -0.06$ eV). At $E = -0.08$ eV, there are no resonant states on $SnO_2$ and the transport across the FTJ is expected to occur via direct tunneling.

These expectations are confirmed by the computed transmission $T$ as a function of energy $E$ across the $SrRuO_3/BaTi(Sn)O_3/SrRuO_3$ FTJ with $SnO_2$ placed at $L = 3$ and ferroelectric polarization pointing to the right. As seen from Fig. 3 (c), there is an onset of transmission at $E = -0.08$ eV so that $T$ increases by more than five orders in magnitude with the increasing energy up to $E_F$. This onset is due to the transition to a resonant tunneling regime. The upward trend results from the increasing number of quantum-well states contributing to transmission at a higher energy, as seen from insets in Fig 3.

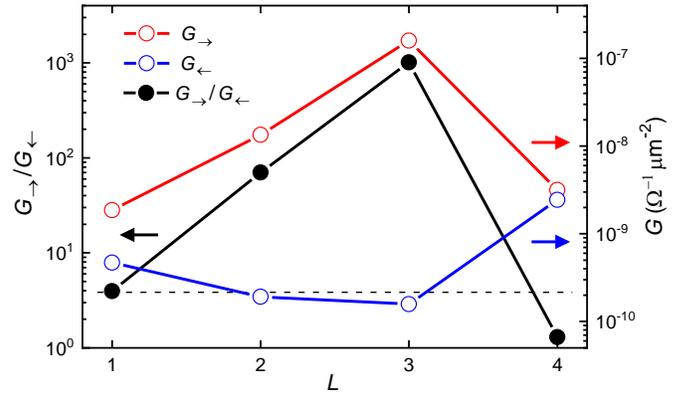

FIG. 4. Conductance $G$ per lateral area of $SrRuO_3/BaTi(Sn)O_3/SrRuO_3$ FTJs as a function of $SnO_2$ layer position $L$ in $BaTiO_3$ for polarization pointing right $G_\rightarrow$ (red dots) and left $G_\leftarrow$ (blue dots) and corresponding ON/OFF conductance ratio $G_\rightarrow/G_\leftarrow$ (black dots). The dashed line shows the junction conductance without the $SnO_2$ layer.

Fig. 4 shows the calculated conductance per area $G = \frac{2e^2}{h}\frac{T}{A}$ (where $T$ is transmission and $A$ is a lateral unit-cell area of the junction) of the FTJs with different positions of the $SnO_3$ layer for two polarization orientations. When polarization is pointing left, conductance $G_\leftarrow$ does not change much as the $SnO_2$ layer is moved from the middle of the barrier to the right interface (blue dots in Fig. 4). This is because the $SnO_2$ LDOS is positioned well above $E_F$ (Figs. 2 (e-h)), and $SnO_2$ acts as a normal barrier independent of its location. In this case, the transport mechanism is controlled by direct tunneling. However, when polarization is pointing right, conductance $G_\rightarrow$ increases fast when the $SnO_2$ layer is moved toward the interface from $L = 1$ to $L = 3$ and then drops down when the $SnO_2$ layer is placed at the interface, $L = 4$ (red dots in Fig. 4). The increase in $G_\rightarrow$ is due to the band bending which pulls the $SnO_2$ LDOS down, reducing the tunneling barrier height. When $SnO_2$ is placed at $L = 3$, the $SnO_2$ LDOS crosses $E_F$, thus providing quantum-well states for resonant tunneling. Due to being separated from the $SrRuO_3$ electrode by an interfacial $BaTiO_3$ barrier layer, these resonant states strongly enhance conductance. On the contrary, when SnO$_2$ is placed at the interfacial layer $L = 4$, due to being in contact with conducting SrRuO$_3$, the SnO$_2$ effectively serves as the termination of the metal electrode. In this case, the quantum well vanishes, electron transport is controlled by direct tunneling, and conductance $G_\rightarrow$ drops down.

To obtain more insight into the mechanism of resonant tunneling, we calculate $\mathbf{k}_\parallel$-resolved transmission $T(\mathbf{k}_\parallel)$ across FTJs. As seen from Fig. 5, independent of the SnO$_2$ position and polarization orientation, $T(\mathbf{k}_\parallel)$ exhibits a cross feature centered at the $\bar{\Gamma}$ point in the 2D Brillouin zone. This feature is intrinsic to BaTiO$_3$ whose $\mathbf{k}_\parallel$-dependent evanescent states reveal the lowest decay rates along the four $\bar{\Gamma} - \bar{M}$ directions [42] and reflect direct tunneling across BaTiO$_3$. For polarization pointing left, we observe qualitatively similar $T(\mathbf{k}_\parallel)$ patterns controlled by the evanescent states of BaTiO$_3$ and independent of the SnO$_2$ location (Figs. 5 (e-h)). This behavior signifies the direct mechanism of tunneling.

On the contrary, for polarization pointing right, $T(\mathbf{k}_\parallel)$ exhibits a "hot spot" around the $\bar{\Gamma}$ point with the transmission magnitude strongly dependent on the SnO$_2$ location $L$ (Figs. 5 (a-d)). The hot spot is most pronounced at $L = 3$ (Fig. 5 (c)), where $T$ is enhanced by four orders in magnitude near the $\bar{\Gamma}$ point. Zooming in on the dashed-line square of Fig. 5 (c) reveals a ring feature in $T(\mathbf{k}_\parallel)$ around the $\bar{\Gamma}$ point (left panel in Fig. 5 (c)) reminiscent to that in the spectral density (Fig. 3 (a)). This correspondence between the $T(\mathbf{k}_\parallel)$ and SD indicates that it is the SnO$_2$ quantum-well states which are responsible for the enhanced conductance $G_\rightarrow$ at $L = 3$ due to resonant tunneling assisted by these states.

When SnO$_2$ is shifted to $L = 2$, i.e. closer to the middle of the barrier, the intensity of the hot spot is reduced (Fig. 5 (b)). In this case, the SnO$_2$ quantum-well band is shifted up in energy (Fig. 2 (b)) and the resonant transmission is featured by the $T(\mathbf{k}_\parallel)$ distribution peaked at the $\bar{\Gamma}$ point (left panel in Fig. 5 (b)). The hot spot vanishes at $L = 1$ (Fig. 5 (a)) due to the SnO$_2$ band being pulled above the Fermi energy (Fig. 2 (a)). When the SnO$_2$ layer is at the interface ($L = 4$), it becomes the termination of the metal electrode. While there is an enhanced transmission near the $\bar{\Gamma}$ point (Fig. 5 (d)), it has the distinctive cross feature reflecting the evanescent states in BaTiO$_3$.

We conclude therefore that both the polarization orientation and the placement of the SnO$_2$ layer in the FTJ control the transport mechanism and conductance. While for polarization pointing left, the transport is governed by direct tunneling independent of the SnO$_2$ position, for polarization pointing right, it strongly depends on the location of the SnO$_2$ layer. In the latter case, when SnO$_2$ is placed at the second ($L = 3$) or third ($L = 2$) $B$O$_2$ layer from the interface, the conductance is strongly enhanced due to resonant tunneling. The resulting ON/OFF conductance ratio reaches a factor of $10^3$ (black symbols in Fig. 4) and could be enhanced even further by proper engineering of the FTJ [43]. These DFT results are corroborated by a simple quantum-mechanical model of tunneling across a potential barrier which contains a quantum well [39].

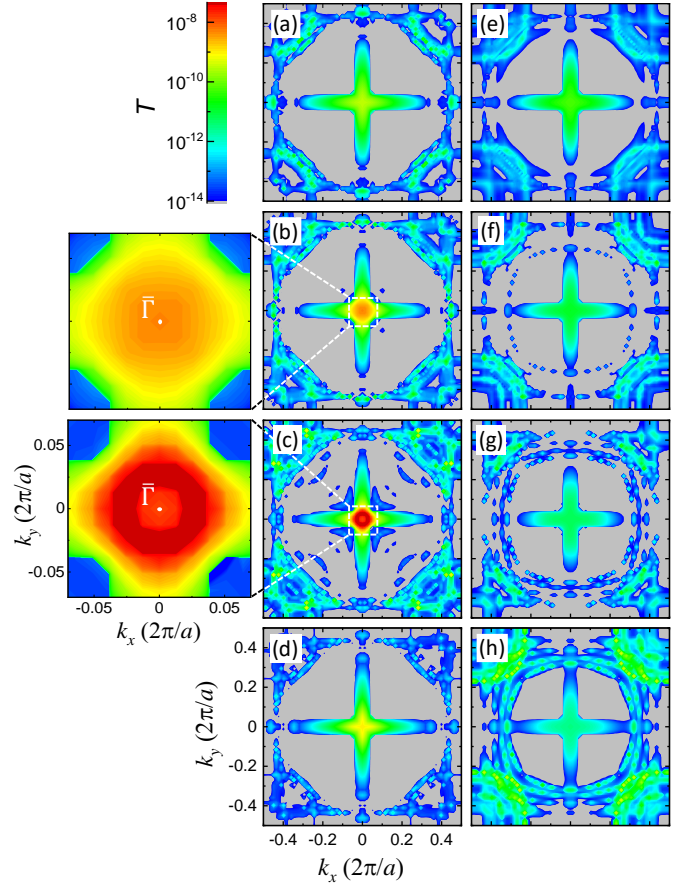

FIG. 5. $\mathbf{k}_\parallel$-resolved transmission across SrRuO$_3$/BaTi(Sn)O$_3$/SrRuO$_3$ tunnel junctions with the SnO$_2$ layer at four different positions $L = 1$ (a, e), $L = 2$ (b, f), $L = 3$ (c, g), and $L = 4$ (d, h) for polarization pointing to the right (a-d) and left (e-h). Left panels in (b) and (c) show $T(\mathbf{k}_\parallel)$ zoomed in around the $\bar{\Gamma}$ point ($k_\parallel = 0$).

In summary, we have proposed that resonant band engineering can serve as viable tool to control the mechanism of conductance and enhance the TER effect in FTJs. For a prototypical SrRuO$_3$/BaTiO$_3$/SrRuO$_3$ FTJ, we demonstrated that a single BaSnO$_3$ layer in the BaTiO$_3$ barrier could form quantum-well states supporting resonant tunneling. The effect is dependent on polarization orientation serving as a switch between resonant and direct tunneling and resulting in a giant TER effect. The proposed approach can be further elaborated to design FTJs with required performance by proper engineering of barrier, resonant-band, and electrode materials and can be exploited to control spin polarization of the tunneling current. The predicted phenomenon is relevant to a ferroelectric field-effect transistor utilizing a BaSnO$_3$ channel, which is expected to exhibit a relatively high mobility [44], and can be helpful in realizing a ferroelectric resonant tunneling diode [45, 46]. We hope therefore that our theoretical predictions will stimulate experimental studies of FTJs and related functional oxide heterostructures with enhanced performance driven by the designed electronic bands.

The authors thank Ding-Fu Shao for valuable discussions. This work was supported by the National Natural Science Foundation of the People's Republic of China (Grants 11974211 and 11974212), Qilu Young Scholar Program of Shandong University, and Taishan Scholarship of Shandong Province. Z.W. acknowledges support from the Taishan Scholar Program of Shandong Province (tsqn201812045). E.Y.T acknowledges support from NSF MRSEC (Grant DMR-1420645).

* liuxiaohui@sdu.edu.cn
** tsymbal@unl.edu

results in an ON/OFF conductance ratio which can reach a factor of $10^7$.

# SUPPLEMENTAL MATERIAL

# Resonant Band Engineering of Ferroelectric Tunnel Junctions

Jing Su[1], Xingwen Zheng[1], Zheng Wen[2], Tao Li[3], Shijie Xie[1], Karin M. Rabe[4],

Xiaohui Liu[1], and Evgeny Y. Tsymbal[5]

[1] School of Physics, State Key Laboratory of Crystal Materials, Shandong University, Ji'nan 250100, China

[2] College of Physics and Center for Marine Observation and Communications, Qingdao University, Qingdao 266071, China

[3] Center for Spintronics and Quantum Systems, State Key Laboratory for Mechanical Behavior of Materials,

Department of Materials Science and Engineering, Xi'an Jiaotong University, Xi'an, Shanxi 710049, China

[4] Department of Physics and Astronomy, Rutgers University, Piscataway, New Jersey 08854, USA

[5] Department of Physics and Astronomy, University of Nebraska, Lincoln, Nebraska 68588-0299, USA

## A. Calculation Methods

First-principles density functional theory (DFT) calculations are performed using the plane-wave pseudopotential code Quantum ESPRESSO [1]. In the calculations, the exchange and correlation effects are treated within the generalized gradient approximation (GGA). The electron wave functions are expanded in a plane-wave basis set limited by a cut-off energy of 550 eV. Periodic boundary conditions are employed for a supercell constructed of 8.5 unit cells of $BaTiO_3$ with one atomic layer of $TiO_2$ substituted by $SnO_2$ and 6.5 unit cells of $SrRuO_3$ (see Fig. 1 in the main text). The in-plane lattice constant of the supercell is constrained to the calculated lattice constant of cubic $SrTiO_3$, $a$ = 3.94 Å. Full internal relaxations are performed to determine the supercell lattice constant and the internal atomic configurations, using the force convergent criterion of 10 meV/Å. For the structural relaxation, the Brillouin zone is sampled with a 6×6×1 Monkhorst-Pack $k$-point mesh.

Transmission $T$ is calculated using a general scattering formalism [2,3] implemented in Quantum-ESPRESSO. The above $SrRuO_3/BaTi(Sn)O_3/SrRuO_3$ supercells are used as scattering regions attached on both sides to semi-infinite $SrRuO_3$ leads. Such FTJ structure has periodic boundary conditions in the $x$-$y$ plane, as described by the transverse Bloch wave vector $\boldsymbol{k}_\parallel = (k_x, k_y)$. $T$ is calculated using a uniform 80×80 $k_\parallel$ mesh in the two-dimensional (2D) Brillouin zone.

## B. Band Alignment between $BaTiO_3$ and $BaSnO_3$

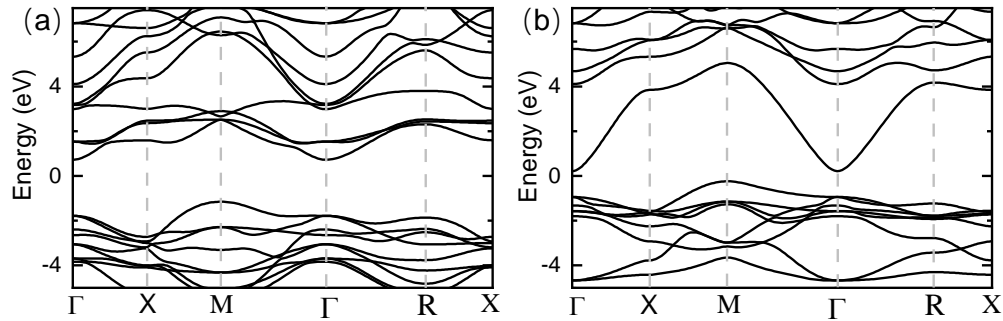

**FIG. S1.** Bulk band structure of strained $BaTiO_3$ (a) and $BaSnO_3$ (b).

Fig. S1 shows the calculated band structure of bulk $BaTiO_3$ and $BaSnO_3$ with the lattice constant of both in the (001) plane being strained to be $a = 3.94$ Å, which is the calculated lattice constant of $SrTiO_3$. Under this constraint, the calculated direct bands gap of $BaTiO_3$ and $BaSnO_3$ at the Γ point are found to be 2.5 eV and 1.2 eV, respectively.

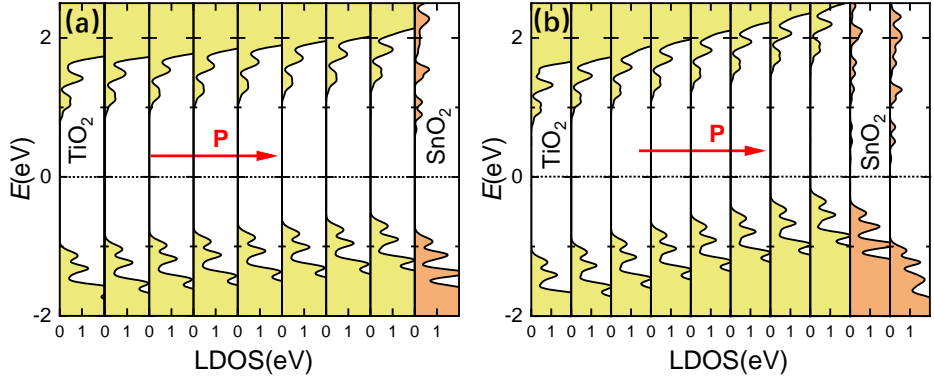

**FIG. S2.** Local densities of state (LDOS) as a function of energy $E$ on $BO_2$ ($B$ = Ti, Sn) atomic layers of the $(BaTiO_3)_n/(BaSnO_3)_m$ superlattice with $n = 8$ and $m = 1$ (a) and $m = 2$ (b).

The smaller band gap of $BaSnO_3$ allows using Sn substitution for Ti in $BaTiO_3$ to form quantum-well states in the band gap of $BaTiO_3$ and thus use them for resonant-band engineering of FTJs. This is evident from our DFT calculations of superlattice heterostructures $(BaTiO_3)_n/(BaSnO_3)_m$, where Sn is substituted for Ti in a monolayer ($m = 1$) or two monolayers ($m = 2$) of (001) $BaTiO_3$. Fig. S2 shows the calculated local densities of states (LDOS) on $BO_2$ ($B$ = Ti, Sn) atomic layers across the heterostructure. It seen that in both cases of $m = 1$ (Fig. S2 (a)) and $m = 2$ (Fig. S2 (b)), the conduction band minimum (CBM) at the $SnO_2$ layer is shifted below the CBM at the $TiO_2$ layers. The conduction band offset increases with the increasing $BaSnO_3$ thickness and produces quantum-well states in the band gap of $BaTiO_3$. It is notable from Figs. S2 (a, b) that these quantum-well states reveal themselves as peaks in the LDOS near the CBM due to the confinement effect.

We note that in Figs. S2 (a,b), the $TiO_2$ LDOS exhibits a slope where the CBM and the valence band maximum (VBM) increase in energy along the direction of polarization. This is due to the bulk polarization of $BaTiO_3$ (40 μC/cm$^2$) being *smaller* than that of $BaSnO_3$ (50 μC/cm$^2$) under the constraint of the $SrTiO_3$ in-plane lattice constant. As a result, the electrostatic field in the $(BaTiO_3)_n/(BaSnO_3)_m$ heterostructure is pointing *parallel* to the polarization within the $BaTiO_3$ layer and but *antiparallel* to the polarization within the $BaSnO_3$ layer.

## C. Resonant Tunneling across an FTJ with Two Layers of $BaSnO_3$

Here we explore the effect of $BaSnO_3$ layer thickness on resonant tunneling and TER by considering a $SrRuO_3/BaTi(Sn)O_3/SrRuO_3$ FTJ with two atomic layers of $BaSnO_3$ inserted in the barrier. DFT calculations of the electronic structure and transmission are performed as described in sec. A. Fig. S3 (a) shows the atomic displacements across the FTJ for polarization pointing to left and right (blue and red curves in Fig. S3 (a), respectively). Similar to the FTJ with one inserted monolayer of BaSnO$_3$ (Fig. 1 in the main text), we observe an enhancement of polar displacement at the SnO$_2$ layers. The calculated LDOS on $BO_2$ ($B$ = Ti, Sn) atomic layers for the two polarization states is shown in Figs. S3 (c, d). It is seen that the LDOS also exhibits features similar to those for the FTJ with one inserted atomic layer of BaSnO$_3$ (Figs. 2 (b, f) in the main text). The major difference, however, is a deeper shift of the SnO$_2$ quantum well-states into the band gap of BaTiO$_3$, which is consistent with the LDOS of the (BaTiO$_3$)$_n$/(BaSnO$_3$)$_m$ superlattice shown in Fig. S2 (b). Also, similar to Fig. S2 (b), the confinement of the BaSnO$_3$ bands is reflected in the splitting of the LDOS near the CBM.

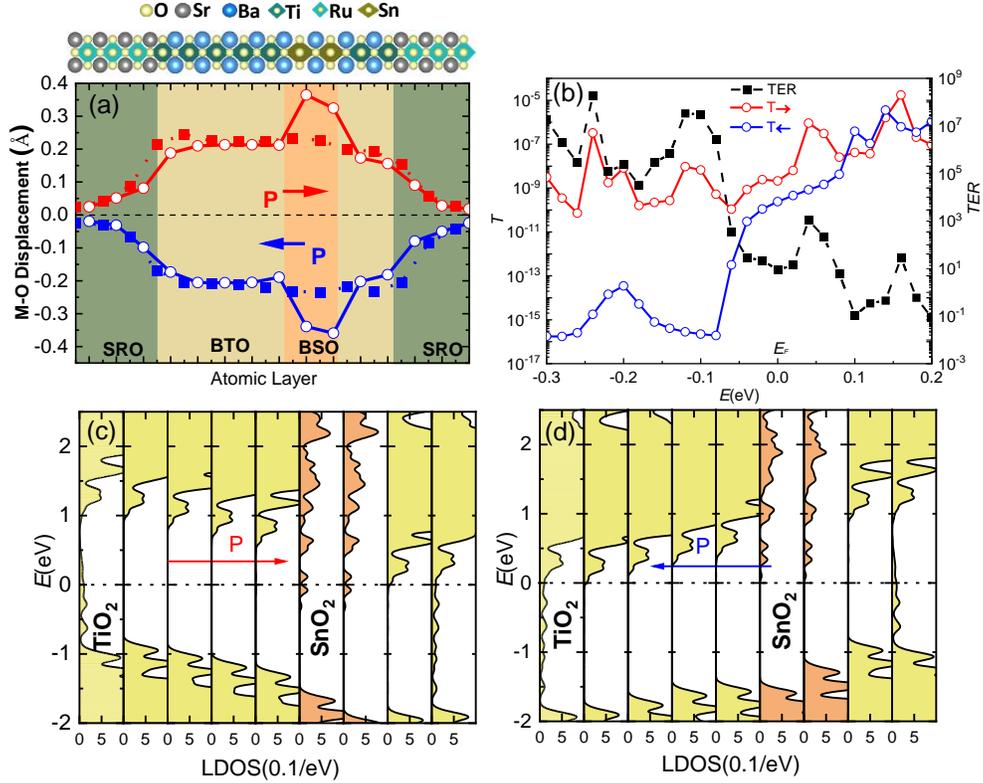

**FIG. S3.** (a) Calculated relative polar displacement between cation (M) and anion (O) on each $AO$ ($A$ = Ba or Sr) and $BO_2$ ($B$ = Ti, Sn, or Ru) atomic layer across the SrRuO$_3$/BaTi(Sn)O$_3$/SrRuO$_3$ supercell (top panel) with two layers of SnO$_2$ substituted for TiO$_2$ in the BaTiO$_3$ barrier. (b) Transmission $T$ per unit-cell area of the FTJ for polarization pointing right (left) $T_\rightarrow$ ($T_\leftarrow$) and the TER ratio $T_\rightarrow/T_\leftarrow$ as a function of electron energy $E$. The Fermi energy $E_F$ is at zero. (c, d) Local densities of states (LDOS) as a function of energy $E$ on each $BO_2$ ($B$ = Ti, Sn) atomic layer of the composite barrier for polarization pointing to the right (c) and left (d). Horizontal dashed lines indicate the Fermi energy.

Comparing Figs. S3 (c) and (d) reveals that polarization switching changes the energy position of the SnO$_2$ LDOS with respect to the Fermi level. The CBM of the SnO$_2$ layer is shifted down in energy for polarization pointing to the right (Fig. S3 (c)), as compared to that for polarization pointing to the left (Fig. S3 (d)). While for the former, the Fermi level crosses the conduction bands of the SnO$_2$ layer, for the latter, the Fermi level lies nearly at the CBM.

Fig. S3 (b) shows the calculated transmission for polarization pointing right (left) $T_\rightarrow$ ($T_\leftarrow$) and the TER ratio

$T_\rightarrow/T_\leftarrow$ as a function of electron energy $E$. It is seen that for polarization pointing right, the transmission $T_\rightarrow$ is high over the whole range of energies $E$ considered, i.e. −0.3 eV < $E$ < 0.2 eV. This is due to the presence of the resonant states in this energy window (see Fig. S3 (c)). On the contrary, for polarization pointing left, the transmission $T_\leftarrow$ exhibits an onset at $E \approx -0.07$ eV, so that it increases by nearly six orders in magnitude with increasing energy $E$ up to $E_F$. This onset is due to the transition to the resonant tunneling regime similar to that shown in Fig. 3 (c) in the main text. As a result, at $E \approx -0.07$ eV, there is a crossover between high and low TER values. As seen from Fig. S3 (b), while for $E < -0.07$ eV, the TER ratio varies in the range of $10^5 - 10^8$, for $E > -0.07$ eV, it is reduced to $1 - 10^3$.

## D. Modeling of Resonant Tunneling

The predicted phenomenon is further elaborated using a simple free-electron model. Within this model, we consider an FTJ with a ferroelectric barrier layer, which contains a thin dielectric of a smaller band gap, placed between two semi-infinite metal electrodes. The electrodes are described using free-electron bands, the Fermi energy of 3.0 eV with respect to the bottom of the bands, and the Thomas-Fermi screening length of 1 Å. The composite tunneling barrier has overall thickness $d = 3.5$ nm, where the thickness of the dielectric layer is set to be 0.4 or 0.8 nm. The ferroelectric and dielectric layers are assumed to have the potential barrier heights of 0.8 eV and 0.2 eV, respectively, with respect to the Fermi energy. The electron effective mass $m^*$ in the barrier region is assumed to be 2 times free-electron mass $m_0$. The polarization of the ferroelectric and dielectric layers are set to be 40 μC/cm² and 50 μC/cm², respectively, as follows from our first-principles calculations using the Born effective charge method. The relative electric permittivity of the barrier is assumed to be $\varepsilon = 90$.

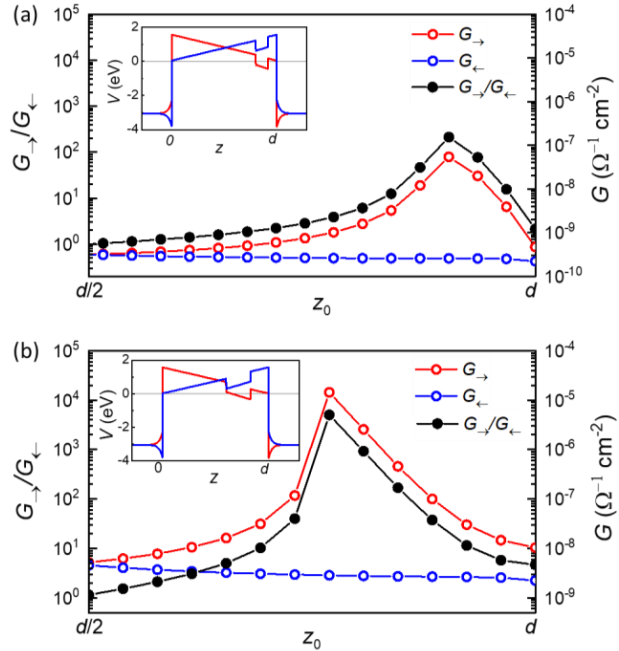

**FIG. S4.** Conductance per area $G$ as a function of the position $z_0$ of the dielectric layer of thickness 0.4 nm (a) and 0.8 nm (b) in the barrier of an FTJ for polarization pointing right $G_\rightarrow$ (red dots) and left $G_\leftarrow$ (blue dots) and the corresponding ON/OFF conductance ratio $G_\rightarrow/G_\leftarrow$ (black dots). Total thickness of the composite barrier is $d = 3.5$ nm. Insets show the potential profiles for the largest values of $G_\rightarrow/G_\leftarrow$.

Using these parameters, we obtain the electrostatic potential profile $V(z)$ across the whole FTJ using the Thomas-Fermi model (insets in Figs. S4 (a) and (b)). Then, the transmission $T$ is calculated using the transfer matrix method. A proper numerical integration over the transverse wave vectors $\mathbf{k}_\parallel$ is performed to calculate the conductance per area $G$ of the FTJ.

Fig. S4 shows results for the calculated conductance as a function of the position $z_0$ of the dielectric layer in the tunneling barrier. It is seen that, for the polarization pointing left (blue dots in the figure), the conductance $G_\leftarrow$ does not change much when the dielectric layer is moved from the middle of the barrier ($z_0 = d/2$) to the right interface ($z_0 = d$). However, for the polarization pointing right (red dots in the figure), the conductance $G_\rightarrow$ reveals a pronounced maximum for a certain position of the dielectric layer, resulting from the resonant tunneling effect. The ON/OFF conductance ratio $G_\rightarrow/G_\leftarrow$ exhibits a maximum at the same position of the dielectric layer (black dots in Fig. S4). As seen from the inserts of Fig. S4, this maximum corresponds to the Fermi energy crossing of the quantum well, which is formed in the barrier by the dielectric layer. The results of this simple model are in a qualitative agreement with our first-principles calculations, further indicating that proper band engineering of an FTJ can used to produce a giant TER effect due to resonant tunneling reversible by ferroelectric polarization.